# How to orient cells in micro-cavities for high resolution imaging of cytokinesis and lumen formation.

Alka Bhat,[1-5] Linjie Lu,[1-5*] Chen-Ho Wang,[6*] Simon Lo Vecchio,[1-5] Riccardo Maraspini,[6] Alf Honigmann,[6] Daniel Riveline[1-5]

*Affiliations:*

[1]Laboratory of Cell Physics ISIS/IGBMC, CNRS and University of Strasbourg, Strasbourg, France
[2]Institut de Génétique et de Biologie Moléculaire et Cellulaire, Illkirch, France
[3]Centre National de la Recherche Scientifique, UMR7104, Illkirch, France
[4]Institut National de la Santé et de la Recherche Médicale, U964, Illkirch, France
[5]Université de Strasbourg, Illkirch, France
[6] Max Planck Institute of Molecular Cell Biology and Genetics, Dresden, Germany

\* Equal contributions

Correspondence to Alf Honigmann and Daniel Riveline

*Abstract:* Imaging dynamics of cellular morphogenesis with high spatial-temporal resolution in 3D is challenging, due to the low spatial resolution along the optical axis and photo-toxicity. However, some cellular structures are planar and hence 2D imaging should be sufficient, provided that the structure of interest can be oriented with respect to the optical axis of the microscope. Here, we report a 3D microfabrication method which positions and orients cell divisions very close to the microscope coverglass. We use this approach to study cytokinesis in fission yeasts and polarization to lumen formation in mammalian epithelial cells. We show that this method improves spatial resolution on range of common microscopies, including super-resolution STED. Altogether, this method could shed new lights on self-organization phenomena in single cells and 3D cell culture systems.

Keywords: Biological Physics, Microfabrication, Super-resolution STED, Cytokinesis, Cell Polarization, Organoids.

# Introduction

Cells have been cultured traditionally on Petri dishes (Petri, 1887) where they can grow easily. In addition, adherent cell culture facilitates microscopy observation within single or a couple of focal planes. This 2D approach has become the standard method to culture and to observe cells and multicellular systems, in basic research, in drug discovery, and in medicine for diagnosis.

However, this method is not optimal for many reasons (Abbott, 2003) . Cells are artificially 'forced' to spread on planar and flat surfaces, potentially inducing artificial shapes and pathological signaling pathways compared to environments encountered in physiological conditions (Weaver, Petersen, Wang, Larabell, Briand, Damsky et al., 1997) (Anders, Hansen, Ding, Rauen, Bissell, Korn, et al., 2003). Cells evolve more physiologically in 3D environments, and classical Petri dishes may not reproduce faithfully this condition. However, 3D cell culture methods, in which cells are typically embedded in hydrogels, pose new challenges to our standard microscope observation techniques. In 3D and sometime also in 2D cell culture many structures/organelles will be ill-oriented with respect to the plane of observation for cells. Placing these organelles within the focal plane of objectives could lead to gain in resolutions in space and in time. These limitations have generated the need to design 3D structures closer to physiological conditions.

In this context, microfabrication has allowed to prepare 3D environments which are compatible with cell growth and organ designs (Sontheimer-Phelps, Hassell, & Ingber, 2019) (Eyckmans & Chen, 2017). Shapes in 3D are generated with micrometer resolution using protocols which can be performed with simple training (see Figure 1). We reported since 2009 (Riveline & Buguin, 2009) (Wollrab & Riveline, 2012) (Riveline, 2012) (Wollrab, Caballero, Thiagarajan, & Riveline, 2016) the preparation of cavities, where cells can be placed vertically with respect to the coverglass. In these 3D configurations, cell division occurs also in a direction perpendicular to the long axis of the cell, but structures such as cytokinetic rings can be seen with unprecedented resolutions. In standard coverslips, rings are perpendicular to the plane of acquisition of objectives: each plane of image capture gives two points, and full rings visualization requires multiple z-stacks acquisitions and 3D reconstruction. In contrast, in these cavities, cytokinetic rings can be seen 'en face' in a single focal plane of acquisition (see Figure 2). Any close to planar cellular structures with such orientations can in principle be imaged with this device and this method.

In this paper, we describe our different procedures and strategies to orient cell structures with 3D microfabrication. We also go one step further and prepare 3D cavities, which are optimized for super-resolution microscopy. We benchmark our system by imaging the cytokinetic rings in fission yeast and in mammalian cells, as well as the polarization and lumen formation in mammalian

MDCK cells after the first cell division. The gain in resolution due to the positioning close to the coverglass and the orientation along the optical axis is significant in time and in space (see Figure 3). For example, clusters of myosin could be resolved with this method with resolution of 200nm with conventional confocal microscopy. They were visualized either by inverted microscope or by having cells imaged directly from below with no interface (Figure 3B). Through-hole cavities allowed us the usage of super-resolution STED microscopy for cytokinetic rings and lumen (Figure 3B.ii and Figure 4) with high NA objectives and minimal optical aberrations. The gain in temporal resolution is also important since organelles can be imaged within sub-seconds in their planes of evolution (see Figure 3C). For example, we reported thanks to this method that myosin clusters were still during cytokinesis in the cell framework. This unexpected observation and its analysis and comparisons with theory and other systems led to the conclusions that myosin clusters generate stress which is associated to the ring constriction in mammalian cells. This supported the notion that myosin clusters dynamics and their self-organizations can serve as generic readouts in cells and in model systems to track and predict changes in shapes (Wollrab, Thiagarajan, Wald, Kruse, & Riveline, 2016). Altogether, these results show that cells inserted vertically in cavity can resolve spatial and temporal cues not accessible so far.

We report below the steps for preparation of 3D micro-patterns by taking the following outline. We start by describing the design and preparation of masks (I), and continue by the fabrication of motifs (II and III). We then explain how to place cells within cavities (V). Troubleshooting is reported in part VI.

## I. Design of mask for optimized cavity preparation.

Masks will serve as support to expose and protect resins from light during the process. In their designs, motifs and inter-distances need to be carefully rationalized.

For each cellular system, cavity dimensions need to be matched to cell dimensions. The first step consists of measuring the distribution of cell diameter right after trypsinization for mammalian cells, and plating cells on a coverslip coated with fibronectin (typically 10µg/ml for 1 hour at room temperature). Distributions for cell diameter can be very sharp like for fission yeasts, and broader for mammalian cells. The mean value is a good starting point for selecting the disk dimension to be prepared on masks.

In addition to cavity diameter, the inter-distance between motifs is important. Inter-distance determines number of cavities per sample. In addition, as reported below, low inter-distance secures a satisfactory filling percentage of cells in cavities. However, this inter-distance should not be too low in order to secure firm attachment of the elastomer stamp. A rule of a thumb consists in taking inter-distance at least twice the cavity size.

After the designed motifs are printed and ready on a photomask, microfabrication process of the motifs follows. The mask prepared in our case had diameters of 17µm through 22µm with an inter-distance of 30µm for mammalian cells or 5µm for fission yeast. Photomasks can be designed using software such as Clewin (Freeware) or Autocad®. Disks are plain or transparent depending on the type of photoresist used in the next steps.

## II.  Microfabrication of motifs on Si wafer.

After the photomask is ready, a light sensitive epoxy-based material (photoresist) is used to obtain motifs on Si wafer. The photoresist (SU-8 in our case) is spin-coated on the wafer with specific coating time and speed to obtain the target height for cavities. In this part, we report SU-8 pillars which are fabricated on Si wafers.

The average diameters for HeLa and MDCK cells were measured to be ~20µm and ~15µm respectively. Therefore, the chosen diameter/height in our case were 22µm/25µm and 17µm/12µm for Hela cells and MDCK cells respectively. The preparation of cavities for fission yeasts is reported in this reference (Riveline, 2012) with special conical shapes to secure firm anchorage of cells.

### Materials
- Silicon wafers
- Wafer tweezers
- Chromium/plastic photomask (with printed motifs to microfabricate)

- SU-8 photoresist (2010 and 2025)
- SU-8 developer
- Acetone
- 2-propanol
- Ethanol

**Equipment**
- Mask aligner (as UV source)
- Spin-coater
- Hot plates (65 °C and 95 °C)
- Disposable graduated dropper (for the removal of bubbles)

**Steps for photolithography:**
1. To ensure that the silicon wafer is contaminant free, it is first cleaned with acetone. It is then dried with a nitrogen stream and followed by similar treatment with ethanol and drying process. Solvent cleaning ensures complete removal of oils and organic residues from the wafer surface. After solvent cleaning, it is advisable to heat up the wafers at 200°C for 10-20 minutes for complete removal of humidity from the surface of wafer.
2. To obtain the targeted height, pour the photoresist (SU-8 2025 for 25µm, and SU-8 2010 for 10µm) on the wafer and remove air bubbles with the help of a plastic dropper. Set a two-step spinning process on the spin-coater. Spin at 500rpm for 10s with an acceleration of 100 rpm/s during the first step to homogenize the layer over the wafer for attaining both 25µm and 12µm height. For a height of 25µm, spin the wafer containing resin at 3000rpm for 30s with acceleration of 300 rpm/s for the second step (*WS-650-23 Spin Coater; Laurell Technologies corporation*, centrifuge and rotor details kept throughout for microfabrication). For a height of 12µm, spin the wafer containing the resin (SU-8 2010) at 3500rpm for 30s with acceleration of 1000rpm/s. This should result in a homogeneous photoresist layer of the required thickness.
3. After obtaining the photoresist evenly spread, pre-bake has to be done to ensure a firm attachment of the photoresist to the wafer. This step has to be performed for 5min at 95°C.
4. After pre-bake, make a firm contact between the wafer and the mask containing the respective motifs. A defective contact will lead to ill-defined motifs. After tight contact with the mask, cross-link the photoresist by exposing the wafer to UV irradiation with a dose of 150 mJ/cm².

The time of exposure will depend on the power of the device (e.g. mask aligner) for the UV light exposure (see Figure 1A.i).

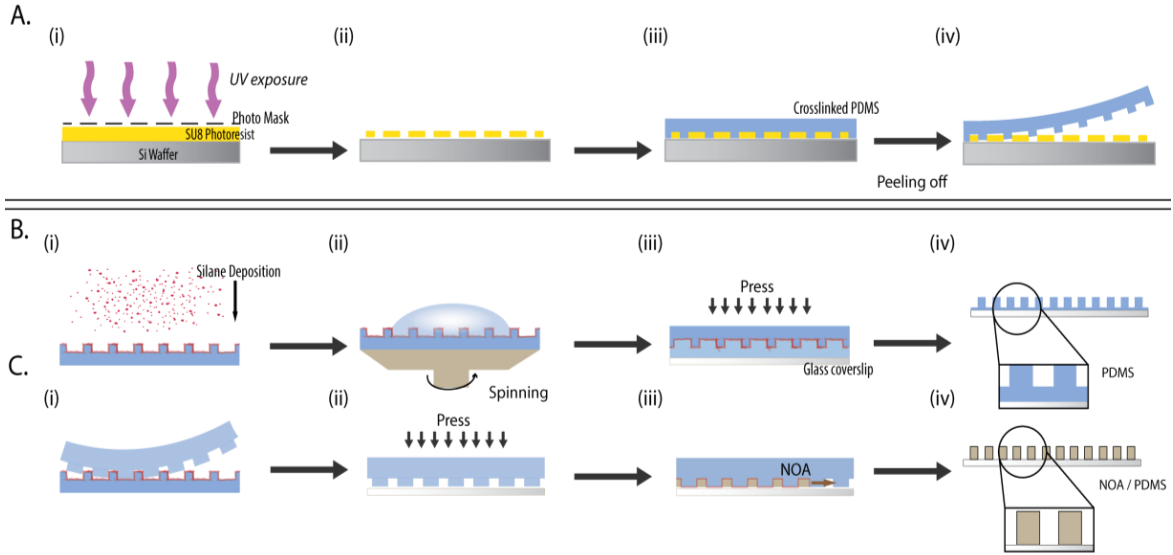

*Figure 1: Microfabrication and preparation of cavities. A. Microfabrication of a silicon master with photoresist SU-8 (A.; (i)) to obtain SU-8 pillars (A.; (ii)). Un-crosslinked PDMS is poured over the silicon master (A.; (iii)) and peeled off to give PDMS cavities (A.; (iv)). B. PDMS cavity preparation with polymer (PDMS) layer at the bottom by first silanizing the PDMS cavities with TMCS (B.; (i)) followed by spin-coating a drop of PDMS (B.; (ii)) to obtain uniform thin layer and crosslinking at 65°C O/N. The assembly is plasma-bonded with a glass coverslip and pressed uniformly (B.; (iii)). Crosslinked PDMS is peeled off from the assembly revealing cavities with PMDS layer at the bottom (B.; (iv)). C. PDMS cavities without polymer layer at the bottom is prepared using UV crosslinking polymer NOA-74 by first using the silanized PDMS master 1 (B.; (i)) with holes and pouring non crosslinked polymer over it, and the sample is allowed to polymerize at 65°C O/N. After polymerization, the PDMS mold with pillars is peeled off from the assembly (C; (i)) and plasma bonded to a glass coverslip (C.; (ii)). The assembly is exposed to TMCS for silanization. Next, the assembly is incubated with NOA-74 (C.; (iii)) and polymerized under UV-A light for minimum 4 hours. PDMS mold with pillars is peeled off carefully from the glass coverslip revealing through-hole cavities (C.; (iv)). Same process can be used to obtain both PDMS and NOA through-hole cavities.*

5. As soon as the exposure is over, follow a post-bake of 1min at 65°C and then 5min at 95°C. An image of the mask should appear on the SU-8 photoresist coating.
6. After post-bake, develop the structures by immersing them in the SU-8 developer solution, while gently agitating the container for around 4min. This will strip off the non-crosslinked resin, leaving behind the motifs. Finally, rinse the surface with 2-propanol to remove the left-over photoresist.

7. White left-over seen over the wafers during 2-propanol treatment is an indicator for residual un-crosslinked resin. Repeat step 6 until rest of the non-crosslinked resin is stripped off the wafer completely.
8. Rinse with 2-propanol and dry the wafer containing motifs using a nitrogen air stream.
9. After the silicon wafers are ready with desired motifs, do a 'hard bake' step at ~150°C for 10 minutes. This is useful for annealing any surface cracks that may have appeared after development; this step is relevant to all layer thicknesses. Structures fabricated in such a way have excellent thermal and mechanical stability (see Figure 1A.ii).
10. After obtaining the motifs, cover the wafers with PDMS (crosslinker: base [1:9]) (see Figure 1A.iii).
11. Remove all air bubbles through desiccation, and finally cure it overnight at 65°C on a leveled surface.
12. Once the elastomer is cured, peel off the PDMS block containing the desired cylindrical pillars with appropriate height (see Figure 1A.iv).

## III. Process for cavity preparation

Cavities are prepared using polymers such as PDMS (heat-crosslinking polymer) or NOA-74 (UV-crosslinking polymer). Different steps are taken to achieve cavities with a thin polymer layer at the bottom, as well as through-hole cavities with no bottom polymer layer.

### Materials
- NOA-74 (Norland products Inc.)
- PDMS (Base and crosslinker; Sylgard 184 Silicone Elastomer)
- Tweezers
- UV-A chamber (for polymer crosslinking)
- Oven (at 65 °C)
- Desiccator
- Sonicator
- 70% ethanol solution

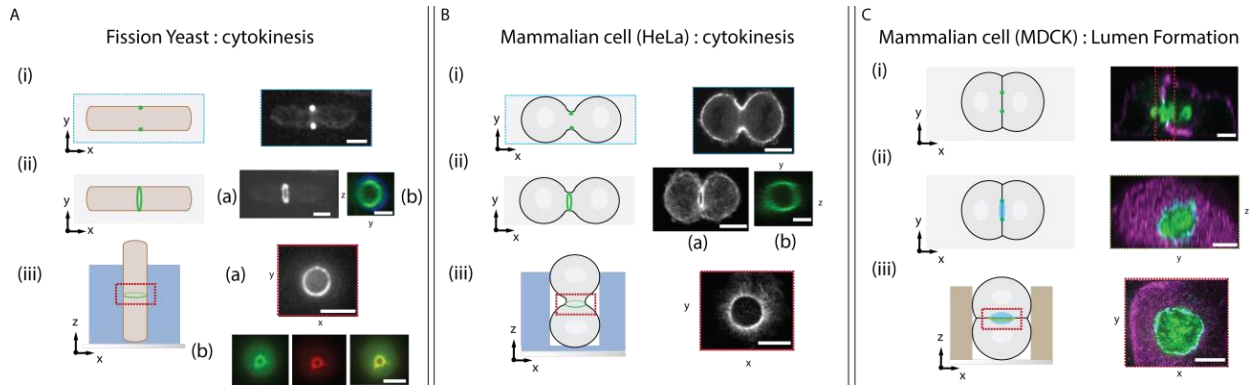

*Figure 2: Orienting cells lead to optimal resolutions in space for a variety of cellular systems. A. Fission yeast cytokinetic ring (tagged-myosin regulatory light chain RLC) visualized on a flat surface (glass coverslip), single plane imaging shows two points in the 'xy' plane corresponding to the ring (A. (i); scale bar 3µm). Reconstitution of ring with 'z' planes projections showing blurred tilted ring (A. (i); scale bar 3µm). Reconstitution of 'yz' planes, (A. (iib); scale bar 3µm) showing a blurred ring and with sectional cut 'en face' (green: pxl1 red for the ring: blankophor for the cell wall). Vertical cells (A. (iiia); scale bar 3µm) allow to reveal myosin clusters around the ring perimeter (Riveline, 2009). Also new structures appear such as arms, and concentric rings (A. (iiib); scale bar 3µm) (green: Pxl1 and red: Rlc1). B. Mammalian cell (HeLa) cytokinetic ring (tagged-myosin heavy chain MHC) visualized on a flat surface (B. (i); scale bar 10µm) with 'xy' plane showing two point accumulations. Reconstitution of 'z' planes projections showing blurred tilted ring (B. (iia); scale bar 10µm). Reconstitution of 'yz' planes, (B. (iib); green: myosin; scale bar 5µm) showing a deformed blurred ring with maximal projection and with sectional cut 'en face'. To visualize the full ring perimeter, cells are oriented vertically with respect to the plane of visualization inside cavities (B. (iii); scale bar 5µm), revealing myosin clusters around the ring perimeter. C. Lumen formation in MDCK doublets, cultured on glass coverslip (with Matrigel) after 24h visualized by confocal microscopy in single plane, showing two points accumulation (light blue:ZO1; lumen). Reconstitution of 'yz' planes showing a blurred and deformed lumen contour with sectional cut 'en face' for MDCK doublets. (C.; (ii)). Gain in information as well as in resolution is achieved after seeding cells inside through-hole NOA-74 cavities showing resolved and undistorted lumen contour (light blue: ZO1, magenta: E-cadherin, green: podocalyxin) Scale bars 3µm.*

## (i) Cavities with polymer layer at the bottom:

PDMS base and the curing agent are mixed in a ratio of 9:1 (v/v). This liquid PDMS is poured over the silicon master obtained via photolithography (for HeLa and MDCK cells), and silicon master is obtained via deep reactive ion etching (for fission yeast, (Riveline, 2012).

1. Liquid PDMS poured over the silicon wafers is degassed (removing trapped air inside the liquid polymer) and cured at 65°C overnight.

2. After curing by heat, the PDMS layer is peeled off (see Figure 1A.iv) from the wafer surface. This peeled off layer is considered as Master 1 with holes in the PDMS block. This is used as replica mold to fabricate pillars used as final motifs for the preparation of cavity.
3. Next, PDMS block with pillars is plasma activated and subsequently passivated with chlorotrimethyl silane (TMCS) under a desiccator for 10min.
4. Small amount of uncured liquid PDMS (~50µl) is poured over the motifs with subsequent spin-coating (and vacuum to hold the PDMS block) with 500rpm for 10s (to homogeneously distribute the liquid PDMS drop) and 1700rpm for 45s to obtain the preferred height (~ 40µm) (see Figure 1B.ii). (*KW-4A Spin coater; SPI supplies*, centrifuge and rotor details kept throughout for Section III (i)). This thickness will be corresponding to the polymer layer at the bottom of glass coverslip.
5. This combination of uncured polymer over PDMS block (containing pillars) is then incubated at 65°C overnight for curing.
6. Cured PDMS block and glass coverslip #0 (100µm thickness) are activated under oxygen plasma. In addition, PDMS block is bonded with the glass coverslip (coated side down). This configuration is pressed for several seconds to ensure tight bonding between them (see Figure 1B.iii).
7. This coverslip-PDMS 'sandwich' is incubated at 65°C overnight. Longer exposure to heat helps the bonding to be stronger between PDMS and coverslip.
8. Next, coverslip is held on both sides of the PDMS block, while peeling it off carefully. This avoids the risk of coverslip breakage during the process.
9. Thin bottom layer of PDMS on the coverslip (~ 40µm) with cavities on top is revealed (see Figure 1B.iv) and Figure 3A.i).

Even though cells are oriented in the right orientation with a good match between cell and cavity dimensions, an elastomer layer at the bottom decreases resolution. This leads to loss in fluorescence as well as in spatial resolution.

A typical example is shown with cytokinetic ring of HeLa cells observed using such cavities with polymer layer at the bottom and an inverted setup: the ring is blurred (see Figure 3A.iia-b). This blurring effect is due to distortion of light caused by this thin PDMS layer at the bottom. Therefore, for such configurations, it is better to visualize these structures with an upright microscope (see

Figure 3A.iiia-b) and this allowed to reveal regularly spaced (0.8µm) myosin clusters around the cytokinetic ring (Wollrab et al., 2016). Visualizing cells in such configuration does not only increase *spatial* resolution, but also *temporal* resolution for capturing fast dynamics (e.g. for cytokinesis in HeLa cells; see Figure 3C.i-ii).

To be able to visualize structures with higher resolutions, cavities without any polymer layer at the bottom (see Figure 1C.iv and Figure 3B.i) are fabricated and cells are visualized using STED super-resolution microscopy (see Figure 3B.iia-b).

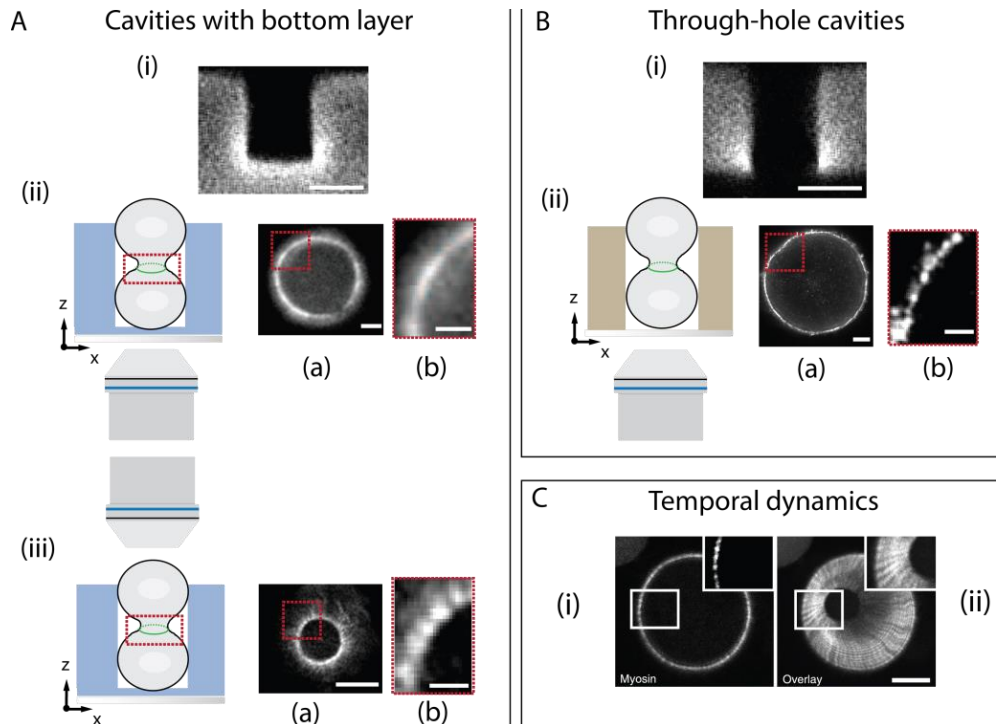

*Figure 3: Gain in spatial and temporal resolutions in cavities; an example with myosin clusters of cytokinetic ring. A. PDMS cavities with polymer (PDMS) layer at the bottom (A; (i) Scale bar 10 µm) is used. Synchronized cells are seeded inside cavities and visualized with an inverted set-up (A. (ii)) showing a blurred cytokinetic ring (tagged-MHC) (A. (iia); scale bar 5µm); inset (red dashed lines) zoomed view showing a blurred portion of the ring (A. (iib); scale bar 3µm; Wollrab, 2014). To better resolve the cytokinetic ring inside a cavity with a PDMS layer at the bottom, imaging is done with an upright microscope (A. (iii)) avoiding the bottom polymer layer, revealing evenly spaced myosin clusters around the cytokinetic ring perimeter (A. (iiia); scale bar 5µm); inset (iib; red dashed lines; scale bar 1µm) zoomed view revealing the distance between clusters to be ~0.8µm. For further gain in resolution and optimized set-up usage, NOA-74 cavities without polymer layer at the bottom (B. (i) Scale bar 10 µm) are prepared. Synchronized cells are seeded inside cavities and visualized with an inverted setup (B. (ii)) using STED microscopy. Visualization of cytokinetic ring revealing individual myosin clusters (B. (iia); scale bar 5µm); inset (iib; red dashed lines; scale bar 1µm) zoomed view revealing myosin cluster inter-distance to be ~0.2 µm. C. Temporal resolution in cytokinetic ring (cluster) dynamics achieved by following the ring closure over time at high frame rate (10s), overlay of 25 frames, (C. (i)) shows clusters moving radially (still clusters) throughout the ring closure (C. (ii)) scale bar 5µm) (Wollrab, 2016)*

**(ii) Cavities without polymer layer at the bottom.**

Cells directly 'touching' the glass surface - without any polymer layer at the bottom - have a significant advantage to gain in resolutions during image acquisition. In addition, this configuration also provides higher freedom of objective working distance as well as the ease in using both upright and inverted microscope setups. Through-hole cavities are achieved with both polymers mentioned earlier, PDMS and NOA-74.

1. Steps from section III (i) steps 1 and 2 are repeated to obtain Master 1 (PDMS block with holes).
2. To obtain PDMS pillars (final motifs for cavity preparation), Master 1 is activated under oxygen plasma. Next, it is passivated with chlorotrimethyl silane (TMCS) under a desiccator for 20-30 mins (see Figure 1B.i). Post silanization PDMS is poured over the Master 1 (with holes), degassed and cured at 65°C overnight.
3. After curing, PDMS is peeled off the Master 1 (PDMS with holes) revealing PDMS block with pillars of target height.
4. To obtain cavities without PDMS layer at the bottom, first, blocks of PDMS with pillars (0.5cm X 0.5cm) are cut and sonicated (5min) with 70% ethanol to remove any dirt and residual silane.
5. These PDMS blocks (with pillars) are used for the preparation of cavities for fission yeast cells, as well as through-hole NOA-74 cavities for HeLa and MDCK cells.
6. Surfaces of PDMS pillars are activated under oxygen plasma, along with the glass coverslip on which pillars are to be bonded.
7. Activated PDMS block with pillars is now turned (activated side down) and put on top of the activated coverslip. The assembly is carefully pressed to ensure proper bonding. Inadequate boding would lead to non through-hole cavities (see Section VI and Figure 5D). Excessive strong bond will result in blocked cavities (see Section VI and Figure 5C). Care is taken that the pressure does not tilt these pillars (see Section VI, Figure 1C.ii and Figure 5A).
8. The assembly is then incubated for passivation with TMCS in a desiccator for 20-30 mins. Passivation is done to prevent irreversible bonding of PDMS pillars with the introduced polymer (NOA-74).

9. Post-passivation: the polymer (NOA-74) is introduced as a drop on one side of the assembly. Capillary action pulls the polymer into the assembly (see Figure 1C.iii). Speed of polymer flow throughout the assembly is dependent on height of the pillars as well as viscosity of the polymer. For example, NOA-74 is less viscous than PDMS and will 'travel' through the assembly quicker and more efficiently than PDMS. Desiccator can be used for 5-7 min for efficient and homogeneous filling of pillars, removing any air bubble trapped inside the assembly.

10. Polymer NOA-74 incubated assembly is then placed for curing (crosslinking) inside a chamber with UV-A source for minimum 4-6 hours.

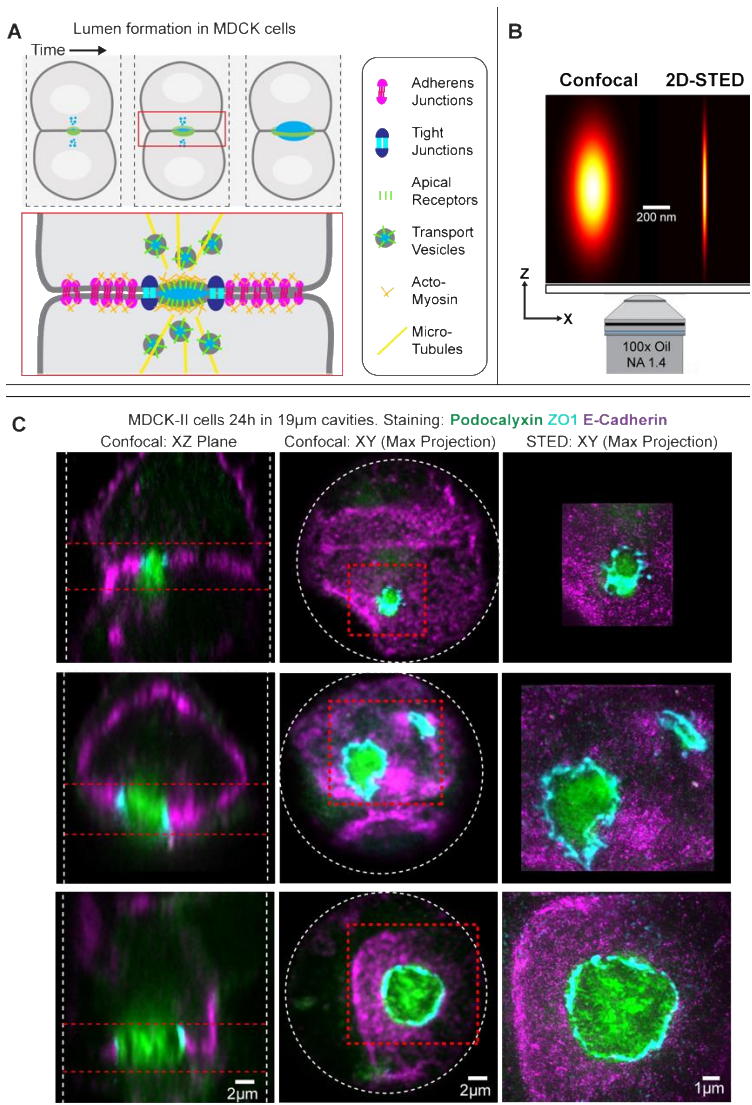

*Figure 4: Visualisation 'en face' of MDCK lumen at the two cell stages in cavities with STED microscopy. A. Sketch of the lumen initiation over time. Zoom shows the cell-cell adhesion interface and some of the key molecular components involved in organising the lumen. (B; (i)) Point spread function of confocal and 2D-STED, showing the 10 fold increase in XY resolution for STED but axial resolution is low. (B; (i)) Illustration of MDCK doublets inside through-hole NOA-74 cavities with lumen (inset; red). (C) Confocal and STED images of MDCK doublets cultured in through-hole NOA-74 cavities for 24h. Cells, which divided in the axial direction, allow to resolve the cell-cell adhesion interface with super-resolution at different stages of lumen expansion. Note that E-cadherin adhesion clusters are clearly resolved in lateral domain in the STED image. Additionally, the apical membrane protein podocalyxin appears to form protrusions in the STED image, which is consistent with micro-villi formation on the apical membrane.*

11. After curing, pillars are carefully peeled off the surface of coverslip. Through-hole NOA-74 cavities are revealed, directly attached to the glass coverslip without any polymer layer at the bottom (see Figure 1C.iv and Figure 3B.i).

The same procedures 9-11 can be used to prepare PDMS cavities as well with no bottom layers. A typical example of cells inside such configuration, imaged with an inverted STED super-resolution microscope, is shown where myosin clusters can be observed at higher resolution (see Figure 3B.iia-b) revealing myosin clusters spaced at ~0.2μm intervals. Another example of oriented structures such as 'lumen' can also be visualized inside similar (PDMS) cavities without polymer layer at the bottom. (see Figure1C (iv) and Figure 4C; STED)

## IV. Choosing the optimal polymer for the process: advantages and drawbacks of different polymers

As described above, two types of polymers, PDMS and NOA-74 are used. There are advantages and drawbacks for each polymer type. For imaging cells inside cavities, it is important to use a biocompatible polymers (Sia & Whitesides, 2003) (Masters, Engl, Weng, Arasi, Gauthier, Viasnoff, 2012).

Also, polymer viscosity can be critical to prepare cavities with small heights. Such configuration would require shorter PDMS pillars, plasma-bonded to coverslip. Incubation of this assembly with a comparatively lower viscosity polymer (80-95cP; NOA-74) is much easier than a polymer with higher viscosity (3500cP; PDMS). If a high viscosity polymer is required, polymer flooding with desiccation is needed. This helps to secure the process with an even distribution of the polymer.

Next, to obtain optimal resolution, the cured polymer with refractive index closer to glass coverslip (~ 1.52) must be chosen. This minimizes distortion of light due to passage through the polymer layer. Refractive indices of both PDMS (~1.41) and NOA-74 (~1.52) are close to glass, with NOA-74 being a better candidate.

Cured PDMS polymer forms strong attachment with glass coverslip. Such property is preferred when robust and long-lasting and/or permanent attachment with coverslip are required. If the polymer layer is to be detached from the surface, polymers with low modulus (NOA-74) are preferred. This allows complete detachment of the layer from the substrate, while keeping its original structure after sonication (10min).

## V. Cell seeding and visualization

We next report the last steps prior observations:

**Mammalian cells (HeLa and MDCK):**

1. Small rectangular piece of PDMS is attached to the coverslip (with cavities) with a drop of un-crosslinked PDMS. This assembly is incubated inside the oven (65°C) for 4h. Such PDMS block will serve as a holder to handle the cavities with help of tweezers in the subsequent steps.
2. Cavities are plasma activated under oxygen plasma.
3. Cavities are further immersed in fibronectin (10 µg/ml) for 4h at room temperature.
4. Cylindrical PMMA piece is introduced inside 50ml tube.
5. Culture media is introduced into the tube until it forms a thin layer over the PMMA piece.

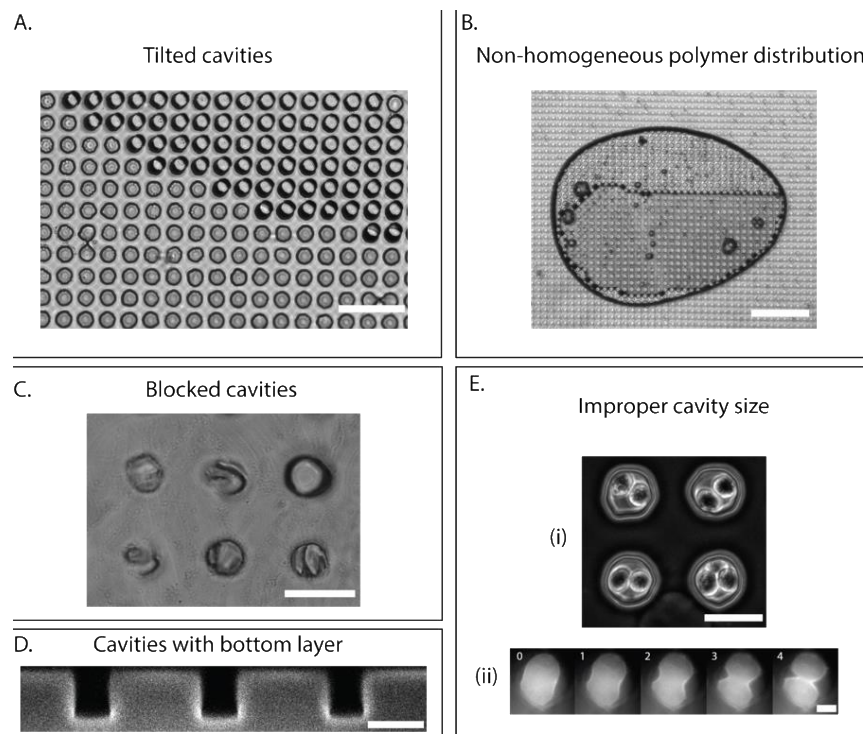

*Figure 5: Issues and troubleshooting.* A. Tilted cavities due to inhomogeneous pressing of PDMS pillars on coverslip prior NOA-74 polymer incubation (scale bar 100µm). B. An air bubble trapped during polymer incubation leads to defects in samples. C. Strong plasma bonding between coverslip and PDMS pillars induces breakage of pillars while peeling them off from cavities. This leads to PDMS blocked NOA-74 cavities (scale bar 40µm). D. Weak plasma bonding between coverslip and PDMS pillars makes the cavities 'non through-hole' (scale bar 20µm). E. Error in matching average cell diameter and cavity dimension, when too large ((i); scale bar 40µm) leads to wrong orientation of dividing cells with respect to the visualization plane ((ii); scale bar 10 µm). (Wollrab, 2014).

6. After the centrifugation steps, cavities are carefully removed and put into a metal holder with selected culture medium, and the sample is ready for acquisition.

## VI.     Troubleshooting for cavity preparation

We list below typical issues during protocols:

- **Tilted PDMS pillars leading to tilted cavities due to pressing during plasma bonding**
  Excessive pressing during plasma bonding (see section III. (ii), step 7.) of coverslip with PDMS pillars can result in tilted pillars (see Figure 5A). Consequently, cells and/or structures under observations would not appear at the correct focus and orientation, leading to poor image resolution.

- **Inhomogeneous cavity distribution due to trapped air bubble**
  In order to achieve homogeneous (NOA-74) prepolymer distribution, one should make sure to introduce the prepolymer on one side of the assembly as a small drop (~10-20μl) and left on the bench for the capillary action to fill the whole assembly. In contrast, polymer introduced from more than one sides could trap air bubble (see section III. (ii), step 9 and Figure 5B).

- **Blocked cavities with PDMS left behind**
  In case of excessive strong bond between coverslip and PDMS pillars (see section III. (ii), step 7.), peeling of PDMS pillars (after NOA-74 crosslinking) may leave PDMS blocked inside the cavities leading to obstructed cavities (see Figure 5C).

- **Polymer layer at the bottom inadequate plasma activation leading to poor bonding**
  Poor plasma bonding between coverslip and PDMS pillars (see section III. (ii), step 7.) could lead to insertion of NOA-74 prepolymer below PDMS pillars leading to non-through-hole cavities (see Figure 5D).

- **Wrong cavity size**

    Cavities should be designed according to the measured, average cell diameter of cells (see Section I). Failing to do so would lead to either horizontal (see Figure 5E.i) or tilted cells (see Figure 5E.ii) inside cavities. This would prevent optimized microscopy and correct orientation of cells and structures under study.

## VII. Conclusions

This article reports strategies to prepare cavities optimized for orientation of cellular structures into the high-resolution plane of microscopes. Many cellular structures could be placed in the 'right' plane with our method. In fact, the usage of 3D environments for studying cells appear to be the natural way to reproduce situations encountered in physiological conditions. The cost is minimal as well as the need for specialized equipment. It is easy to foresee numerous applications of this new point of view in a literal sense to reveal new structures and dynamics with relevance for basic research and for medical applications.

## Acknowledgements:

We thank the Riveline Lab and the Imaging Platform of IGBMC for discussions. This study was supported by the Human Frontier Science Program number RGP0050/2018. A.B. and SLV were also funded by the University of Strasbourg. This work with the reference ANR-10-LABX-0030-INRT has been supported by a French state fund through the Agence Nationale de la Recherche under the frame programme Investissements d'Avenir labelled ANR-10-IDEX-0002-02. C-H.W and R.M. were funded by the Max Planck Society (ELBE fellowship) and by the Deutsche Forschungsgemeinschaft (DFG, German Research Foundation) - Project Number 112927078 - TRR 83.